\documentstyle[12pt]{article}

\begin{document}

\noindent

\def\med{{1\ov 2}}
\def\hepth#1{ {\tt hep-th/#1}}

\def\be{\begin{equation}}
\def\ee{\end{equation}}
\def\bes{\begin{equation*}}
\def\ees{\end{equation*}}

\def\beqa{\begin{eqnarray}}
\def\beqas{\begin{eqnarray*}}
\def\eeqa{\end{eqnarray}}
\def\eeqas{\end{eqnarray*}}
\def\bea{\begin{eqnarray}}
\def\eea{\end{eqnarray}}

\def\cl{\mbox{\tiny (class)}}

\def\etal{{\it et al.\/}}
\def\ie{{\it i.e.\/}}
\def\eg{{\it e.g.\/}}
\def\Cf{{\it Cf.\ }}
\def\cf{{\it cf.\ }}

\def\al{\alpha}
\def\lam{\lambda}
\def\blam{\bar\lambda}
\def\th{\theta}
\def\bth{\bar\theta}
\def\bsigma{\bar\sigma}
\def\bpsi{\bar\psi}

\def\bu{{\bar 1}}
\def\bd{{\bar 2}}
\def\bt{{\bar 3}}
\def\bc{{\bar 4}}
\def\dgam{\dot\gamma}
\def\dal{\dot\alpha}
\def\dbet{{\dot\beta}}

\def\Re{{\rm Re}}
\def\Im{{\rm Im}}

\def\H{{\cal H}}
\def\tr{{\rm tr}}
\def\Tr{{\rm Tr}}
\def\F{{\cal F}}
\def\N{{\cal N}}

\def\d{\partial}
\def\ov{\over}
\def\bD{\bar D}

\def\pder#1#2{{{\partial #1}\over{\partial #2}}}
\def\der#1#2{{{d #1}\over {d #2}}}
\def\ppder#1#2#3{{\partial^2 #1\ov\partial #2\partial #3}}
\def\dpder#1#2{{\partial^2 #1\ov\partial #2 ^2 }}
\def\bemat{\left(\begin{array}}
\def\enmat{\end{array}\right)}
\def\theequation{\thesection.\arabic{equation}}

\def\Fpk{\alpha\!\cdot\!\!\F'_k}
\def\Fpu{\alpha\!\cdot\!\!\F'_1}
\def\Fpb{\beta\!\cdot\!\!\F'_1}
\def\Fpd{\alpha\!\cdot\!\!\F'_2}
\def\Fppk{\alpha\!\cdot\!\!\F''_k\!\!\cdot\! \alpha}
\def\Fppu{\alpha\!\cdot\!\!\F''_1\!\!\cdot\! \alpha}
\def\Fppb{\beta\!\cdot\!\!\F''_1\!\!\cdot\! \beta}
\def\Fppd{\alpha\!\cdot\!\!\F''_2\!\!\cdot\! \alpha}
\def\cdotsh{\!\cdot}

\begin{titlepage}
\begin{flushright}
{ ~}\vskip -1in
US-FT/6-98\\
YCTP-P11-98\\
\hepth{9805172}\\
May 1998\\
\end{flushright}

\vspace*{20pt}
\bigskip

\centerline{\Large WHITHAM HIERARCHIES, INSTANTON CORRECTIONS}
\bigskip
\centerline{\Large AND SOFT SUPERSYMMETRY BREAKING}
\bigskip
\centerline{\Large IN ${\cal N}=2$ $SU(N)$ SUPER YANG-MILLS THEORY}
\vskip 0.9truecm
\centerline{\large\sc Jos\'e D. Edelstein$^{\,a,}$\footnote{\tt
edels@fpaxp1.usc.es}, Marcos Mari\~no$^{\,b,}$\footnote{\tt
marino@waldzell.physics.yale.edu} and
Javier Mas$^{\,a,}$\footnote{\tt jamas@fpaxp1.usc.es}}

\vspace{1pc}

\begin{center}
{\em

 $^a$ Departamento de F\'\i sica de Part\'\i culas, Universidade de Santiago de
Compostela,\\
E-15706 Santiago de Compostela, Spain.\\

\bigskip
$^b$ Department of Physics, Yale University\\
New Haven, CT 06520,
 USA.} \\

\vspace{5pc}

{\large \bf Abstract}

\end{center}

We study ${\cal N}=2$ super Yang-Mills theory with gauge group $SU(N)$ from the
point of view of the Whitham hierarchy.
We develop a new recursive method to compute the whole instanton
expansion of the prepotential using the theta function associated to the root
lattice of the group. Explicit results for the one and
two-instanton corrections in $SU(N)$ are presented.
Interpreting the slow times of the hierarchy as additional
couplings, we show how they can be promoted to spurion superfields
that softly break $\N=2$ supersymmetry down to $\N=0$.
This provides a family of nonsupersymmetric deformations of the theory,
associated to higher Casimir operators of the gauge group. The $SU(3)$ theory
is analyzed in some detail.

\end{titlepage}


\section{Introduction and Conclusion}

Among
the large number of toy-models that have been proposed in the literature,
in the aim to capture the essentials of non-perturbative QCD, the one solved
exactly by Seiberg and Witten \cite{SeiWitt} stands out as a
breakthrough. Apart from its unquestionable beauty, this work is remarkable in
that it condenses and gives shape to the most  beautiful ideas and conjectures
about Yang-Mills theories that have been suggested over the last 30 years like,
for instance, duality or quark confinement by monopole condensation.

 Certainly, as a toy model, the Seiberg-Witten solution only represents an
approximation to the real world: in order to get an exact answer, the price to
pay is the need for ${\cal N}=2$ supersymmetry. The generalization of the
initial breakthrough from $SU(2)$ to $SU(N)$ was soon unraveled
 \cite{sun}. Another step towards the real
world was to avoid supersymmetry. Since in the Seiberg-Witten ansatz
supersymmetry was
an essential ingredient from the very beginning, the strategy was to break it
softly, trying to preserve the analytic properties of the solution.
The spurion formalism \cite{spurion} proved to be instrumental, and
results were reported for the first time in \cite{soft}, in the context of
$SU(2)$ with and without additional matter. These results were generalized in
\cite{moresoft} to $SU(N)$ and refined in \cite{luisIyII,hsu}.

Seemingly, a totally unrelated topic is that of integrable hierarchies.
In its origin this is a subject related to non-linear
differential equations that appear in classical mechanics of systems with
either
finite or infinite number of degrees of freedom. Although this is a vast
subject
in itself, it is only recently that some unifying language has emerged.
A close relationship
between integrable models and supersymmetric quantum field theory
 was observed sometime ago in the context of two-dimensonal topological
conformal
field theories, obtained by twisting ${\cal N}=2$ superconformal models
\cite{twodi}.

It is by now a well established fact that the Seiberg-Witten solution for the
effective
theory of ${\cal N}=2$ super Yang-Mills can be embedded into the Whitham
hierarchy
associated to
the periodic Toda lattice \cite{toda, todamar}. The link between both
constructions is
summarized in the statement that the prepotential of the ${\cal N}=2$
Yang-Mills theory
corresponds to the logarithm of the quasiclassical tau function. The interplay
between Whitham hierarchies and two-dimensional
superconformal models suggests
understanding the times of the Whitham hierarchy as coupling constants of
composite operators also in the four-dimensional context, and, thereafter, the
prepotential as the generating function of correlation functions for them.
Recently, this interpretation
has proven to be very useful
in understanding some aspects of the twisted version of ${\cal N}=2$ Yang-Mills
theory
\cite{mw,moore1,moore2, moore3, losev,takasaki}. In particular, it has been
shown in \cite{ITEP}
that the slow times of the Whitham hierarchy are the appropriate variables to
understand
the structure of contact terms in the twisted theory.

In this paper the structure of the effective action of ${\cal
N}=2$ theories from the point of view of the underlying
integrable hierarchy will be explored, and we will show that
the results in \cite{ITEP} have
interesting applications for the dynamics of the ${\cal N}=2$, $SU(N)$ theory.
We will see that the integrable structure constraints the semiclassical
expansion of the prepotential in such a way that the knowledge of the
one-loop contribution essentially determines the instanton corrections.  This
provides in fact a new method  to compute the prepotential to any given
instanton number.
The inputs for this computation are the following: first of all, the RG
equation \cite{matone, cobi}, which relates the first derivative of the
prepotential with respect to the quantum scale,
$\partial {\cal F}/\partial \Lambda$, to the quadratic Casimir. The second
ingredient is the equation derived in \cite{losev,ITEP}, which relates
$\partial^2 {\cal F}/\partial \Lambda^2$
to the theta function associated to the root lattice of the gauge group.
These relations allow for a recursive computation, starting from the
one-loop
contribution to the prepotential. Explicit
results for the one and two instanton corrections that agree with those
previously obtained in the literature will be presented. A general formula
for the three instanton correction is also written down.

We will also show that the slow times
of the Whitham hierarchy can be understood as spurion superfields that softly
break
supersymmetry down to ${\cal N}=0$, in the spirit of \cite{soft, moresoft,
luisIyII}. In the original
approach to the soft breaking of ${\cal N}=2$ supersymmetry, the quantum scale
$\Lambda$ is promoted to a spurion superfield, and this generates a series of
terms that explicitly break
supersymmetry. However, these terms are associated to the quadratic Casimir of
the
gauge group, as this Casimir is in essence the dual variable to the quantum
scale. In this
way, the soft breaking to ${\cal N}=0$ using $\Lambda$ as a spurion is the
analog of the
soft breaking to ${\cal N}=1$ using the operator ${\rm Tr}\,\Phi^2$. But one
can
not implement,
in this restricted approach, the ${\cal N}=0$ analog of the ${\cal N}=1$
supersymmetry
breaking operator associated to a higher Casimir operator, like ${\rm Tr}
\,\Phi^3$  for
$SU(3)$.
A natural extension of this formalism is provided by the Whitham
hierarchy. In principle, the variables that appear in the prepotential in the
framework of the Whitham
hierarchy are different from the original variables of the Seiberg-Witten
ansatz, but can be related
in a precise way. In fact, the first
slow time of the hierarchy, $T_1$, can be identified with the quantum scale
$\Lambda$, and the times $T_n$, with $n=2, \dots, N-1$, are dual to particular
homogeneous combinations of the higher Casimir operators. In this
way, the Whitham hierarchy can be interpreted as a family of supersymmetry
breaking
deformations of the original theory associated to the higher Casimir operators
of the gauge group.

The results presented in this paper can be extended in many ways. We have
restricted
ourselves to the theory without matter hypermultiplets and to the gauge group
$SU(N)$.
One could generalize this approach to other gauge groups and/or matter content,
and this would provide a powerful computational tool to obtain instanton
expansions,
along the lines explained in this paper. Another avenue for future research is
the connection with string theory and D-branes. In \cite{evans} it has been
shown that some nonsupersymmetric configurations of branes can be interpreted
as
softly broken ${\cal N}=2$ theories, and on the other hand the approach via
integrable systems has also been casted in the context of D-brane
configurations
\cite{intbrane}. It would be very interesting to explore the connection between
these two
problems, and obtain in this way a new  family of nonsupersymmetric
deformations of
MQCD.

The organization of this paper is the following: in section 2, the
relation between
Whitham hierarchies and the Seiberg-Witten solution is reviewed, following
\cite{ITEP}, and clarify
the relation between the slow time variables and the variables of the
Seiberg-Witten
prepotential. In the remaining sections we present two independent 
applications of the equations  so far. 
In section 3, we study the instanton expansion of the
prepotential in the semiclassical region and explain the new method to compute
this expansion.  Explicit computations
for the one and two-instanton corrections are presented for comparison.
In section 4, we promote the slow times to spurion superfields and we analyze
the
resulting theory once supersymmetry is broken down
to ${\cal N}=0$. Finally, the $SU(3)$ theory is discussed in some detail in
section 5.

\section{Whitham Hierarchies and Seiberg-Witten Ansatz}
\setcounter{equation}{0}
\indent

The low-energy dynamics of $N=2$ super Yang-Mills theory with gauge group
$SU(N)$ is described
by the hyperelliptic curve \cite{sun}
\be
y^2 = P^2(\lambda,u_k) - 4\Lambda^{2N},
\label{hyper}
\ee
where  $P(\lambda,u_k) = \lambda^{N} - \sum_{k=2}^N u_k \lambda^{N-k}$
is the characteristic polynomial of $SU(N)$ and $u_k,~ k=2,...,N$
are the Casimirs of the gauge group. They provide Weyl invariant
coordinates on ${\cal M}_\Lambda$, the quantum moduli space of vacua of the
theory.
This curve
has genus $g=r$, where $r=N-1$ is the rank of the group. As explained
in \cite{toda}, the curve
(\ref{hyper}) can be identified with the spectral curve of the $N$ site
 periodic Toda lattice and, moreover
\cite{toda, todamar}, the prepotential of the effective theory is essentially
the logarithm of the quasiclassical tau function and hence depends on the slow
times of the corresponding Whitham hierarchy. Here the
integrable approach along the lines of \cite{ITEP} will be followed,
and a subfamily of slow times $T_n$ with $1\leq n\leq N-1$ will be considered.
As we mentioned in the introduction, when the Seiberg-Witten ansatz
is embedded into the Whitham hierarchy,  one has to clarify the relation
between the variables and parameters in both approaches.
 Although there is an indication of how this goes in
\cite{ITEP} (see also \cite{Carroll}) we shall pause
here to discuss this point, focusing on $\Lambda$ and $T_1$.

In principle, $\Lambda$ and $T_1$ are different variables.
 $\Lambda$ appears, in the Seiberg-Witten solution, in the hyperelliptic curve
(\ref{hyper}) describing the
moduli space of vacua ${\cal M}_{\Lambda}$. Let
$A^i$ and $B_i$ denote a symplectic basis of homology cycles for this curve,
$i=1, \dots, r$.  The $a^i$ variables of the prepotential, for the duality
frame associated to the cycles $A^i$,
are given by the integrals over these cycles of  a certain meromorphic
one-form:
\be
a^i(u_k,\Lambda) ={1 \over 2\pi i }  \oint_{A^i} {\lambda P'(u_k)\ov
\sqrt{P^2(u_k) -
4 \Lambda^{2N}}
}d\lambda,
\label{laa}
\ee
where $P'= dP/d\lambda$, and the same expression holds for the dual variables
$a_{D,i}$ with $B_i$ instead of
$A^i$.
A dependence of $u_k$ upon $\Lambda$ is induced by solving
$a^i(u_k,\Lambda) =$constant, as $u_k =  f_k(a^j,\Lambda)$.  That is
to say, by using $(a^i,\Lambda)$ as coordinates for the moduli space.

On the other hand, in the context of the Whitham hierarchy, the slow times
$T_n$ appear associated to meromorphic differentials of second kind
$d\hat\Omega_n$ (in the notation of \cite{ITEP})
\beqa
\alpha^i(u_k,T_1,T_2,...) &=&  \sum_{n\geq 1} {T_n \over 2\pi i }\oint_{A^i}
d\hat\Omega_n
\cr
&=&
\sum_{n\geq 1} {T_n \over 2\pi i } \oint_{A^i}{P(u_k)^{n/N}_+ P'(u_k)\ov
\sqrt{P^2(u_k) - 4} }
    d\lambda \cr
&=& T_1 a^i(u_k,1) + {\cal O}(T_{n>1}) ~,
\label{alpha}
\eeqa
where $\left(\sum_{k=-\infty}^\infty c_k\lambda^k\right)_+ =
\sum_{k=0}^\infty c_k\lambda^k$. Also here the $\alpha_{D\,i}$ are defined by
the
same
expression with $B_i$ replacing $A^i$.
  Similarly, we may choose to solve for
$u_k(\alpha^i,T_1,T_2,...)$ by demanding that $\alpha^i$ in (\ref{alpha}) be
independent
of all $T_n$. The induced dependence $u_k = g_k(a^j,T_n)$ solves the Whitham
equations.

The recovery of the Seiberg-Witten solution goes as follows. First we
define the rescaled times $\hat T_n$ and ``vevs" $\hat u_k$:
\be
\hat T_n = T_n/T_1^n, \,\,\,\,\,\,\  \hat u_k = T_1^k u_k,
\ee
with $\hat T_1 = 1$. It is easy to see that (\ref{alpha}) can be written as
\be
\alpha^i(u_k,T_1,\hat T_2,\hat T_3,...) = \sum_{n\geq 1} {\hat T_n \over 2\pi i
}
\oint_{A^i}{P^{n/N}_+(\hat u_k) P'(\hat u_k)\ov \sqrt{P^2(\hat u_k) - 4T_1^{2N}}
}
    d\lambda.
\label{hatalpha}
\ee
In particular, after setting $\hat T_2 = \hat T_3 =...= 0$
we find that
$$
\alpha^i(u_k,T_1,\hat T_{n>1} = 0) = T_1 a^i(u_k,1) =
a^i(\hat u_k,\Lambda=T_1) ~.
\label{alaa}
$$
In conclusion, {\em   we may identify $T_1$ with $\Lambda$ in the submanifold
$\hat T_2 =
\hat T_3 = ...= 0$, provided the moduli space is parametrized with $\hat u_k$
instead of  $u_k$.}

With this correspondence in mind, let us now focus on the 
derivatives of the prepotential $\F(\alpha,T)$. The computation of such
derivatives  from the Whitham hierarchy  was the main result of the paper 
\cite{ITEP}. We just list them
here for completeness:
\beqa
\pder{\F}{T_n} &=& {\beta\ov 2\pi i n} \sum_{m} m T_m\H_{m+1,n+1} ~, \nonumber
\\
\ppder{\F}{\alpha^i}{T^n} &=& {\beta\ov 2\pi i n} \pder{\H_{n+1}}{a^i} ~,
\label{completeness} \\
\ppder{\F}{T_m}{T_n} &=& -{\beta\ov 2\pi i}\left(
\H_{m+1,n+1} + {\beta\ov mn}\pder{\H_{m+1}}{a^i}\pder{\H_{n+1}}{a^i}
{1\ov i\pi}\d_{\tau_{ij}}\log \Theta_E(0|\tau)\right) ~.\nonumber
\eeqa
In these equations, $\Theta_E(0|\tau)$ denotes Riemann's theta function
with a certain characteristic $E$, evaluated at the origin\footnote{
We follow the
convention of \cite{tata} in the definition of the theta function
$\Theta[\vec \alpha, \vec \beta](\xi|\tau)$ associated to the genus $r$
hyperelliptic
curve, and with characteristics $\vec \alpha=(\alpha_1, \dots, \alpha_r)$,
$\vec
\beta=(\beta_1, \dots, \beta_r)$:
\be
\Theta [\vec \alpha, \vec \beta] (\xi|\tau)= \sum_{n_k \in {\bf Z} } \exp
\bigl[ i \pi \tau_{ij}(n_i+ \alpha_i)
(n_j+ \alpha_j) + 2\pi i (n_i+ \alpha_i)(\xi + \beta_i) \bigr].
\label{thetaf}
\ee
Therefore, in this normalization $\d_{\tau_{ij}} = {1\ov 4\pi i} \d^2_{ij}$.
See also \cite{rauch}.
}
; $\beta= 2N,~m,n = 1,...,r=N-1$, and
  derivatives with respect to $T_n$ are taken at constant $\alpha^i$.
The functions $\H_{m,n}$ are certain homogeneous combinations of the
Casimirs $u_k$, given by
$$
\H_{m+1,n+1} = {N\ov mn} \hbox{res}_\infty \left(
P^{m/N}(\lambda)dP_+^{n/N}(\lambda) \right) = \H_{n+1,m+1} ~,
$$
and $$\H_{m+1} \equiv \H_{m+1,2} = {N\ov m} \hbox{res}_\infty P^{m/N}(\lambda)
d\lambda = u_{m+1} + {\cal O}(u_m).$$
Here res$_P$ stands for the usual Cauchy residue at the point $P$.
We have for instance $\H_{2,2}
= \H_{2} = u_2,~\H_{3,2} =\H_{3}= u_3$ and $\H_{3,3} = u_4+ {N-2\over 2N}
u_2^2$.

As they stand, the expressions given in (\ref{completeness}) are not suitable
for
application to the Seiberg-Witten solution. Therefore, and in view of the
previous considerations, we define the following change of variables
\be
\log\Lambda=\log T_1, \,\,\,\,\,\,\ \hat T_n=T_1^{-n} T_n, ~~~~~~(n\ge 2),
\label{cambio}
\ee
and, consequently,
\be
\pder{~}{\log\Lambda} = \sum_{m\geq 1} m T_m \pder{~}{T_m}~, \,\,\,\,\,\,\
\pder{~}{\hat T_n} =  T_1^n \pder{~}{T_n}, \,\,\,\,\ (n\geq 2)
\ee
With the help of these expressions, it is now straightforward to reexpress
all the formulae in (\ref{completeness}) as
derivatives of $\F$ with respect to $\alpha^i, \hat T_n$
and $\Lambda$. Most of the factors $T_1$ can be used to promote
$u_k$ to $\hat u_k$ or, rather, to the homogeneous combinations thereof:
\be
\hat \H_{m+1,n+1} = T_1^{m+n}\H_{m+1,n+1} ~~~~ \Rightarrow ~~~~
\hat \H_{m+1} = T_1^{m+1}\H_{m+1}
\ee
(since $\H_{m+1}= \H_{m+1,2}$).
The remaining $T_1$'s are absorbed in making up
$\hat a^i
\equiv T_1a^i(u_k,1) = a^i(\hat u_k,T_1)$.
Altogether we find
\beqa
\pder{\F}{\log\Lambda\,} & = & {\beta\ov 2\pi i} \sum_{n,m\geq 1}
m\hat T_m\hat T_n\hat\H_{m+1,n+1} ~~, ~~~~~~~~~ \pder{\F}{\hat T_n}
{}~=~ {\beta\ov 2\pi i n} \sum_{m\geq 1} m\hat T_m \hat\H_{m+1,n+1} ~,
\nonumber \\
\ppder{\F}{\alpha^i}{\log \Lambda} & = & {\beta\ov 2\pi i} \sum_{m\geq
1}\hat T_m\pder{\hat\H_{m+1}}{\hat a^i} ~~, ~~~~~~~~~~~~~~~
\ppder{\F}{\alpha^i}{\hat T_n} ~=~ {\beta\ov 2\pi i
n}\pder{\hat\H_{n+1}}{\hat a^i} ~, \nonumber \\
\dpder{\F}{(\log\Lambda)} & = & - {\beta^2\ov 2\pi i}\sum_{m,n\geq 1}
\hat T_m  \hat T_n \pder{\hat\H_{m+1}}{\hat a^i} \pder{\hat\H_{n+1}}{\hat
a^j}\, {1\ov i\pi} \d_{\tau_{ij}}\log\Theta_E(0|\tau) ~, \label{lasecu} \\
\ppder{\F}{\log\Lambda\,}{\hat T_n} & = & - {\beta^2\ov 2\pi i
n}\pder{\hat\H_{n+1}}{\hat a^i} \sum_{m\geq 1}\hat T_m
\pder{\hat\H_{m+1}}{\hat a^j}\, {1\ov i\pi}
\d_{\tau_{ij}}\log\Theta_E(0|\tau) ~, \nonumber \\
\ppder{\F}{\hat T_m}{\hat T_n} & = & - {\beta\ov 2\pi i} \left(
\hat\H_{m+1,n+1}+{\beta\ov mn} \pder{\hat\H_{m+1}}{\hat
a^i}\pder{\hat\H_{n+1}}{\hat a^j}{1\ov i\pi}
\d_{\tau_{ij}}\log\Theta_E(0|\tau) \right) \nonumber ~,
\eeqa
with $m,n\geq 2$. In these expressions $\hat T_1 = 1$.
The restriction to the submanifold
$\hat T_2 = \hat T_3=...=0$ yields formulae which are suited for
the Seiberg-Witten analysis. Notice that in this subspace
$\alpha^i(u_k,T_1,\hat T_{n>1} = 0)=
\hat a^i$, hence
\beqa
\pder{\F}{\log\Lambda\,} & = & {\beta\ov 2\pi i } \hat\H_{2} ~~,~~~~~~~~
{}~~~~~~~~~~~~~~~~~~ \pder{\F}{\hat T_n}
{}~=~ {\beta\ov 2\pi i n}  \hat\H_{n+1} ~, \nonumber \\
\ppder{\F}{\hat a^i}{\log\Lambda} & = & {\beta\ov 2\pi i}\pder{\hat\H_{2}}{\hat
a^i} ~~,~~~~~~~~~~~~~~~~~~~ \ppder{\F}{\hat a^i}{\hat T_n} ~=~ {\beta\ov 2\pi i
n}\pder{\hat\H_{n+1}}{\hat a^i} ~, \nonumber \\
\dpder{\F}{(\log\Lambda)} & = & -{\beta^2\ov 2\pi i}
\pder{\hat\H_{2}}{\hat a^i}\pder{\hat\H_{2}}{\hat a^j}{1\ov i\pi}
\d_{\tau_{ij}}\log\Theta_E(0|\tau) ~, \label{lasecudef} \\
\ppder{\F}{\log\Lambda\,}{\hat T_n} &=& -{\beta^2\ov 2\pi i n}
\pder{\hat\H_{2}}{\hat a^i}\pder{\hat\H_{n+1}}{\hat a^j}{1\ov i\pi}
\d_{\tau_{ij}}\log\Theta_E(0|\tau) ~, \nonumber \\
\ppder{\F}{\hat T_m}{\hat T_n} &=& -{\beta\ov 2\pi i} \left(
\hat\H_{m+1,n+1}+{\beta\ov mn}
\pder{\hat\H_{m+1}}{\hat a^i}\pder{\hat\H_{n+1}}{\hat a^j}{1\ov i\pi}
\d_{\tau_{ij}}\log\Theta_E(0|\tau) \right) ~, \nonumber
\eeqa
with $m,n\geq 2$. As a check, notice that the formula (1.11) in \cite{ITEP}
follows
directly from combining the second and the sixth formulae of the list, \ie
\be
\pder{\hat\H_{m}}{\log\Lambda\,} = -\beta\pder{\hat \H_2}{\hat
a^i}\pder{\hat\H_{m}}{\hat
a^j} {1\ov i\pi}\d_{\tau_{ij}} \log \Theta_E(0|\tau),
\label{qucas}
\ee
and the first equation in (\ref{lasecudef}) is precisely the RG equation
derived in
\cite{matone,cobi} (see below eq.(\ref{mato}).
Hereafter, we will always work with the coordinates (\ref{cambio})
and, therefore, hats will be omitted everywhere.

According to \cite{ITEP}, the characteristic $E$ appearing in
(\ref{lasecu})--(\ref{lasecudef}) is an even, half-integer characteristic
associated to a particular splitting of the roots of the discriminant.
The explicit form of $E$ can be obtained using the connection with the twisted
version of the
${\cal N}=2$ theory investigated in \cite{mw, moore1, moore2, losev,
takasaki}.
The theta function involved in (\ref{lasecu})--(\ref{lasecudef}) is the same
one that appears in the blow-up formula derived in \cite{moore2} and further
discussed in \cite{takasaki} from the point of view of the
Toda--Whitham hierarchy. When no non-abelian magnetic fluxes are turned on, the
blow-up factor of the lattice sum is \cite{moore2}:
\be
 \Theta[\vec \alpha, \vec \beta](t \vec V \vert \tau)=\sum_{n_i } e^{i \pi
\tau_{ij} n_i n_j
   + i t V_i n_i  -  i \pi \sum_in^i},
\label{blowupfactor}
\ee
where $V_i={\partial u_2 \over \partial a^i}$. From here, we read off
\be
\vec \alpha=(0, \dots, 0)
                           ~~~~~\hbox{ and}~~~~~~
\vec \beta = (1/2, \dots, 1/2).
\label{carac}
\ee
This is the characteristic $E$ of the theta function in
(\ref{lasecu})--(\ref{lasecudef}) when we express it  in terms of {\it
electric}
variables. Notice that it is even, half-integer, and for
$SU(2)$ the associated theta function is the Jacobi $\vartheta_4 (z| \tau)$,
in agreement
with the result in Appendix  B of \cite{ITEP}.

We will see in the next section
that the above identification of Seiberg--Witten and Whitham variables,
together with this choice of the characteristic, allows us to find the
appropriate instanton expansion in the semiclassical region.

\section{Instanton Corrections}
\setcounter{equation}{0}
\indent

Instanton calculus provides one of the few non-perturbative links between the
Seiberg-Witten ansatz and the microscopic non-abelian field theory
that it is supposed to describe effectively at low energies. For this reason,
since the very advent of the work of
Seiberg and Witten, there has been a lot of work on this particular topic with
the aim of relating the explicit
non-perturbative computations with the predictions of the exact solutions.  On
the one hand, different techniques have been developed to extract instanton
expansions from the hyperelliptic curves
\cite{matone,klemm,itoyangtwo,dhoker,itoyang, schnitzer}. On the other hand,
instanton corrections have also been explicitely computed
 in the microscopic theory
\cite{instcal,instantones} (see \cite{hunter}
for a review and
a list of references).
In \cite{matone} it was realized that the non perturbative relation
\be
\F - \med a_{D,i} a^i = {\beta\ov 4\pi i}  u
\label{mato}
\ee
is very useful to derive a recursion relation for the instanton contributions.
In order to compute the instanton corrections, however, (\ref{mato}) is not
sufficient and
one needs additional input. This is usually provided by the Picard-Fuchs
equations for the
periods. The Picard-Fuchs equations are difficult to derive and solve when the
rank of the gauge group
is larger than one, although techniques from topological Landau-Ginzburg
theories can make them more instrumental in order to obtain the one-instanton
correction to the prepotential for the ADE series \cite{itoyang}.
The procedure we will use here is rather different. As we will see, it turns
out that the equation
for $\partial^2 {\cal F}/\partial (\log \, \Lambda)^2$ in (2.12), together with
(\ref{mato}),
provides enough information to obtain the instanton expansion of the
prepotential in the semiclassical region to any order, and we don't have to
make use of the Picard-Fuchs equations. We then see that the connection of
$SU(N),~
{\cal N} = 2$ super Yang--Mills theory with Toda--Whitham
hierarchies embodies in a natural way a recursive
procedure to compute all instanton corrections\footnote{When this work was
finished, we noticed that recursive relations for the prepotential were 
recently derived from modular anomaly equations for mass deformed ${\cal N}=4$ 
super Yang-Mills theories in powers of the mass of the adjoint hypermultiplet 
\cite{MNW}. In a particular limit, this theory reproduces the results corresponding
to pure ${\cal N}=2$ super Yang-Mills theory.}. The essential ingredient that
makes this possible is the relation of the derivatives of the prepotential with
the theta function associated to the root lattice of the gauge group.

\subsection{Recursive Procedure from the Prepotential Theory}

To begin with, let us fix our conventions for the expansion of the prepotential
in the semiclassical region.
We choose the basis $H_k = E_{k,k}- E_{k+1,k+1}$  for the Cartan subalgebra
and $E_{k,j},k\neq j$ for the raising and lowering operators.
Let $\{ \alpha_i\}_{i=1,...,r}$ stand for the simple roots of $SU(N)$ and
$(\alpha,\beta)$ denote the usual inner product constructed with the
Cartan-Killing form. The dot product $\alpha\cdotsh\beta \equiv
2(\alpha,\beta)/(\beta,\beta)= (\alpha,\beta^\vee)$.
We have that $\alpha_i\cdot\alpha_j = C_{ij}$, with $C_{ij}$ the Cartan
matrix,
while $\lambda^i\cdotsh\alpha_j = \delta^i{_j}$ define the fundamental weights.
In particular this means that $\alpha_i = \sum_j C_{ij}\lambda^j$.
The simple roots generate the root lattice $\Delta = \{\alpha = n^i
\alpha_i| n^i\in {\bf Z}\}$, and the fundamental weigths its dual, the
weight lattice
$\Delta^\vee$.

The instanton expansion of the prepotential is:
\be
\F = {1\ov 2N}\tau_0 \sum_{\alpha_+} Z_{\alpha_+}^2
+ {i\ov 4 \pi} \sum_{\alpha_+} Z_{\alpha_+}^2 \log \, {Z_{\alpha_+}^2\ov
\Lambda^2}
+{1\ov 2 \pi i}\sum_{k=1}^\infty \F_k(Z)
\Lambda^{2Nk}.
\label{elprep}
\ee
In this expression $a= a^i\alpha_i$ and $Z_\alpha = \alpha\cdotsh a$.
Also, $\alpha_+$ denotes a positive root and $\sum_{\alpha_+}$  a sum over
positive roots. The expansion is in powers of $\Lambda^{\beta}$, where
$\beta=2N$ for
$SU(N)$, and $k$ is the instanton
number. We then have
\be
\dpder{\F}{(\log \Lambda)} = {1\ov 2\pi i}
\sum_{k=1}^\infty (2N k)^2\F_k(Z) \Lambda^{2Nk}~,
\label{makdj}
\ee
which, after (\ref{lasecudef}), should be equated to
\be
\dpder{\F}{(\log\Lambda)\,} =
-{\beta^2\ov 2\pi i} \pder{\H_2}{a^i} \pder{\H_2}{a^j} {1\ov
i\pi}\d_{\tau_{ij}}
\log \Theta_E(0|\tau).
\label{lateta}
\ee
The derivative of the quadratic Casimir also has an  expansion that
can be obtained
from the RG equation and (\ref{elprep}):
\beqa
\pder{\H_2}{a^i} &=& {2\pi i\ov \beta} \ppder{\F}{a^i}{\log\Lambda\,}
\cr &=&
C_{ij} a^j +
 \sum_{k=1}^\infty
 k \F_{k,i} ~ \Lambda^{2Nk}
\cr
&\equiv& \sum_{k=0}^\infty H_i^{(k)} \Lambda^{2Nk}
\label{expanh}
\eeqa
where $\F_{k,i}= \d\F_k/\d a^i$, and use has been made of the fact that ${1\ov
2N} \sum_{\alpha_+}
Z_{\alpha_+}^2 =
{1\ov 2} a^i C_{ij} a^j$. The couplings in the semiclassical region are
obtained again from the expansion (\ref{elprep}):
\beqa
\tau_{ij} &=&\ppder{\F}{a^i}{a^j} \cr
&=& {i\ov 2\pi}
\sum_{\alpha_+} \pder{Z_{\alpha_+}}{a^i} \pder{Z_{\alpha_+}}{a^j}
 \log \left({ Z^2_{\alpha_+} \ov \Lambda^2} \right)
+{1\ov 2\pi i} \sum_{k=1}^\infty \F_{k,ij} \Lambda^{2Nk} ~.
\label{latau}
\eeqa
with $\F_{k,ij}= \ppder{\F_k}{a^i}{a^j}$. For convenience, in (\ref{latau})  a
term ${i\ov 2\pi}\sum_{\alpha_+} \pder{Z_{\alpha_+}}{a^i}
\pder{Z_{\alpha_+}}{a^j}\left( {2 \pi i \ov N}\tau_0 - 3\right) $ has been
set to zero by a suitable adjustment of the bare coupling $2\pi i\tau_0 = 3N$.
Of course, we
may shift $\tau_0$ to any value by appropriately rescaling $\Lambda$, and this
will be
reflected in our choice for the normalization of the $\F_k$. One has to be
careful
with this normalization when comparing our final expressions with similar
computations in the literature.

The term involving the couplings that appear in the theta function $\Theta_E$
is now
\be
i\pi \, n^i\tau_{ij} n^j = \sum_{\alpha_+} \log
\left({Z_\alpha\ov \Lambda}\right)^{-(\alpha\cdot\alpha_+)^2} + \med
\sum_{k=1}^\infty~
(\Fppk) ~\Lambda^{2Nk},
\label{tauroot}
\ee
where $\alpha= n^i\alpha_i$ and
\beqa
\Fppk
 &\equiv& \sum_{i,j}n^i\F_{k,ij} n^j
\cr
&=&
\sum_{\beta, \gamma\in \Delta}(\alpha\cdotsh\beta)
\ppder{\F_k}{Z_{\beta}}{Z_{\gamma}}(\gamma\cdotsh\alpha)~.
\eeqa
The appropriate characteristic $E$ for the theta function $\Theta_E$ in the
semiclassical region has been given in (\ref{carac}). Inserting (\ref{tauroot})
in the theta function, we obtain
\beqa
\Theta_E(0|\tau) &=& \sum_{\vec n} \exp\biggl[ i\pi n^i\tau_{ij} n^j + i \pi
\sum_{k} n_k\biggr]
\cr &=&
\sum_{\alpha\in\Delta}(-1)^{\rho\cdot \alpha}
\prod_{\alpha_+}\left( {Z_{\alpha_+}\ov
\Lambda}\right)^{-(\alpha\cdot\alpha_+)^2}
\prod_{k=1}^\infty\exp\left({\med (\Fppk) \Lambda^{2Nk}}\right)
\cr &=&
\sum_{r=0}^\infty \sum_{\alpha\in\Delta_r} (-1)^{\rho\cdot\alpha}
\prod_{\alpha_+} Z_{\alpha_+}^{-(\alpha\cdot\alpha_+)^2}
\prod_{k=1}^\infty\left(\sum_{m=0}^\infty {1\ov 2^m
m!}\left(\Fppk\right)^m\,
\Lambda^{2Nkm}\right) \Lambda^{2Nr}
\cr &\equiv&
\sum_{p=0}^\infty \Theta^{(p)} \Lambda^{2Np}.
\label{expantheta}
\eeqa
In the previous expression, $\rho = \sum_{i=1}^{N-1} \lambda^i$.
$\Delta_r\subset\Delta$ is a subset of the root lattice composed of those
lattice vectors
$\alpha$ that fulfill the constraint
$\sum_{\alpha_+}(\alpha\cdot\alpha_+)^2
= 2Nr$.
In particular $\Delta_1$ is the root system, {\it i.e.} the simple roots
together with their Weyl reflections.
In other words, as the root system itself, $\Delta_1$,
forms an orbit of the Weyl group,
the one- and two-instanton contributions will come from a sum over
the Weyl orbit of, say, $\alpha_1$. On the other hand $\Delta_r$, for $r>1$,
will be in general a union of Weyl orbits, since Weyl reflections are easily
seen
to be automorphisms of $\Delta_r$. Therefore $\Theta^{(p)}$ is Weyl invariant
by
construction. The first few terms in the expansion (\ref{expantheta}) are given
by
\[ \Theta^{(0)} = 1 ~, ~~~~~~~~~~
\Theta^{(1)} = \sum_{\alpha\in\Delta_1}(-1)^{\rho\cdot\alpha}
\prod_{\alpha^+} Z_{\alpha_+}^{-(\alpha\cdot\alpha_+)^2} ~, \]
\[ \Theta^{(2)} =
\sum_{\alpha\in\Delta_1}(-1)^{\rho\cdot\alpha}\,\med (\Fppu )
\prod_{\alpha^+} Z_{\alpha_+}^{-(\alpha\cdot\alpha_+)^2} 
+ \sum_{\beta\in\Delta_2}(-1)^{\rho\cdot\beta} \prod_{\alpha^+} 
Z_{\alpha_+}^{-(\beta\cdot\alpha_+)^2} ~. \]
In the logarithmic derivative, the theta function appears in the denominator,
and we have the expansion
\be
\Theta(0|\tau)^{-1} = \sum_{l=0}^{\infty} \Xi^{(l)}(\Theta) \,
\Lambda^{2Nl} ~.
\label{tauinv}
\ee
Here $\Xi^{(0)}(\Theta) = 1$ and for $\Xi^{(l)}(\Theta)$ we can write
in general
\be
\Xi^{(l)}(\Theta) = \sum_{(p_1,...,p_k) \in {\bf N}^k}^{p_1 + 2p_2 + ...
+ kp_k = l} \xi_{(p_1,...,p_k)} \prod_{i=1}^l (\Theta^{(i)})^{p_i} ~,
\label{tauinvcoef}
\ee
where the coefficients $\xi$ are parametrized by the
partition elements $(p_1,...,p_k)$. The first few values for these
parameters are, for example,
\[ \xi_{(1)} = -1 ~, ~~~ \xi_{(2,0)} = 1 ~, ~~~ \xi_{(0,1)} = -1 ~, ~~~
\xi_{(3,0,0)} = -1 ~, ~~~ \xi_{(1,1,0)} = 2 ~, ~~~ \xi_{(0,0,1)} = -1 ~, \]
and using these values we can immediately obtain the lower $\Xi^{(l)}(\Theta)$.
Similarly, the derivative of
the theta function with respect to the period matrix is given by
\beqa & & \!\!\!\!\!\!\!\!\!\!\!\!\!\!\!\!\!\!\!\!\!\!\!\!
 {1\ov i\pi}\d_{\tau_{ij}}\Theta_E(0,\tau) = \sum_{ n} n^i n^j
 \exp\biggl[ i\pi n^k\tau_{kl} n^l + i \pi \sum_{k} n_k\biggr]
 \cr
&=& \sum_{r=1}^\infty \sum_{\alpha\in\Delta_r} (-1)^{\rho\cdot\alpha}
(\alpha\cdotsh\lambda^i)(\alpha\cdotsh\lambda^j)
\prod_{\alpha_+} Z_{\alpha_+}^{-(\alpha\cdot\alpha_+)^2}
\prod_{k=1}^\infty\exp\left({\med (\Fppk) \Lambda^{2Nk}}\right)
\Lambda^{2Nr} \cr
&
\equiv& \sum_{p=1}^\infty \Theta_{ij}^{(p)} \Lambda^{2Np} ~.
\label{expanthij}
\eeqa
Now, collecting all the pieces and inserting them back into (\ref{lateta}),
we find for $\F_k(Z)$ the following expression:
\be
\F_k(Z) = - k^{-2}
\sum_{p, q, l=0}^{p+q+l = k-1}\sum_{ij} H_i^{(p)} H_j^{(q)}
\Theta_{ij}^{(k-p-q-l)} \Xi^{(l)} ~,
\label{elresul}
\ee
in terms of the previously defined coefficients.

If we look at the coefficients in (\ref{elresul}),
it is easy to see that the expressions they involve depend on $\F_1,\F_2,...$
up to $\F_{k-1}$.
In fact, although both $H^{(p)}$ and $\Theta^{(p)}$ depend
on $\F_1,....\F_p$, the indices within parenthesis in (\ref{elresul}) reach at
most the value $k-1$ as $\Theta_{ij}^{(0)}=0$. Moreover $\Theta_{ij}^{(k)}$
depends on
$\F_1,...,\F_{k-1}$ since the vector $\alpha=0$ is missing from the lattice
sum.
This fact implies the possibility to build up a recursive procedure
to {\em compute all the instanton coefficients by starting just from  the
perturbative contribution to $\F(a)$ in} (\ref{elprep}).

\subsection{Lower Instanton Corrections}

As we have seen, (\ref{elresul}) gives the instanton expansion of the
prepotential
in the semiclassical region, for ${\cal N} = 2$ super
Yang--Mills theory with gauge group $SU(N)$. We emphasize the fact that the
essential ingredients are summarized in (\ref{lasecudef}). We believe that a
direct comparison
of (\ref{elresul})  with other
computations of $\F_k$ presented in the literature provides an independent
test of the proposal made in \cite{ITEP}. Also, our results show that the
embedding of the Seiberg--Witten theory inside an integrable hierarchy,
aside from its theoretical interest,  provides an alternative device
for some computations in $SU(N)$ ${\cal N} = 2$ super Yang--Mills theory.
Here we shall give a fairly manageable general expression
for the one-, two-, and three-instanton contributions in $SU(N)$,
and compare it with the answers that can be
found in the literature. We would like to point out that, although we have
focused on $SU(N)$, the form of the instanton corrections that we have
presented should be generalizable to other cases. In this respect we point
out that relations such as (\ref{qucas}) hold for all the simply laced 
algebras \cite{losev}.

The one-instanton contribution is given by (\ref{elresul}) with $k=1$. In this
case the
expression is rather simple and reads
\beqa
\F_1 & = &  -
\sum_{ij} H^{(0)}_i H^{(0)}_j \Theta_{ij}^{(1)} \nonumber \\
& = & -
\sum_{\alpha\in\Delta_1}Z_\alpha^2~ (-1)^{\rho\cdot\alpha}\prod_{\alpha^+}
Z_{\alpha^+}^{-(\alpha\cdot\alpha^+)^2} ~,
\label{efe1}
\eeqa
 As pointed out in \cite{klemm},
in general there is not a unique form for the
$\F_k$ when written in terms of the $Z_\alpha$'s, since these are not
independent variables. An unambiguous expression should come out for $\F_k$
when
written in terms of the $a_i$'s or, for example, in terms of symmetric
polynomials
thereof, such as the classical values of the Casimirs.

The two-instanton contribution can also be easily worked out from
(\ref{elresul}),
and turns out to be
\beqa
\F_2 & = & - \frac{1}{4}
\left(  \Theta_{ij}^{(2)} H_i^{(0)} H_j^{(0)}
+ \Theta_{ij}^{(1)}(2 H_i^{(1)}H_j^{(0)} -
H_i^{(0)}H_j^{(0)}\Theta^{(1)})\right)
\nonumber \\
& = & -\frac{1}{4}\left(\sum_{\alpha\in\Delta_1}(-1)^{\rho\cdot\alpha}
\prod_{\alpha_+} Z_{\alpha_+}^{-(\alpha\cdot\alpha_+)^2}
\left[ \F_1 + 2 (\Fpu ) Z_\alpha + \med (\Fppu ) Z_\alpha^2 \right] \right. 
\nonumber \\ 
& & \left. + \sum_{\beta\in\Delta_2}Z_\beta^2~ (-1)^{\rho\cdot\beta}\prod_{\alpha^+}
Z_{\alpha^+}^{-(\beta\cdot\alpha^+)^2} \right) ~,
\label{efe2}
\eeqa
where $\alpha\!\cdot\!\!\F'_k = n^i\F_{k,i} =
\sum_{\beta\in\Delta}(\alpha\!\cdot\beta)\d\F_k/\d Z_{\beta}$.  Furthermore,
our
proposal for
the three-instanton correction gives
\beqa
\F_3 & = & - {1\ov 9}\left(
\sum_{\alpha\in\Delta_1}(-1)^{\rho\cdot\alpha} \prod_{\alpha^+}
Z_{\alpha^+}^{-(\alpha\cdot\alpha^+)^2} \left[\rule{0mm}{5mm} 4\F_2
+ 4 (\Fpd) Z_\alpha + (\Fpu)^2 \right.\right. \nonumber \\
& & \left.\left. + ~\med (\Fppu)\left( \F_1 + 2 (\Fpu) Z_\alpha \right)
+ {1\ov 8} (\Fppu)^2 Z_\alpha^2 + \med (\Fppd) \, Z_\alpha^2 \right]
\right. \nonumber \\
& & \left. + \sum_{\beta\in\Delta_2}(-1)^{\rho\cdot\beta}
\prod_{\alpha_+} Z_{\alpha_+}^{-(\beta\cdot\alpha_+)^2}
\left[ \F_1 + 2 (\Fpb ) Z_\beta + \med (\Fppb ) Z_\beta^2 \right] 
\right. \nonumber \\
& & \left. + \sum_{\gamma\in\Delta_3}(-1)^{\rho\cdot\gamma}
\prod_{\alpha^+} Z_{\alpha^+}^{-(\gamma\cdot\alpha^+)^2}\,
Z_\gamma^2\!\rule{0mm}{7mm}\right),  \nonumber
\eeqa
etc.
The above expressions make patent the recursive character of the procedure.

It is quite cumbersome to show that our formulae
 match the results in the literature for a generic value of $N$.
Nevertheless, we have checked several particular cases using symbolic
computation. For the one-instanton contribution we found the following results
\[ \F_1^{SU(2)} = 2^{-3} \Delta_0^{-1} ~~~ , ~~~
\F_1^{SU(3)} = {3 \ov 2} u_0 \Delta_0^{-1} ~~~ , ~~~
\F_1^{SU(4)} = [8 u_0^3 - 36 v_0^2 + 32 u_0 w_0] \Delta_0^{-1} \]
\beqas
\F_1^{SU(5)} & = & [8 u_0^3 (v_0^2 - 3 u_0 w_0) - 18 v_0^2 (3 v_0^2
- 13 u_0 w_0) - 16 w_0^2 (11 u_0^2 + 20 w_0) \\
& & - 40 v_0 x_0 (u_0^2 - 15 w_0) - 250 u_0 x_0^2] ~\Delta_0^{-1} ~,
\eeqas
where the zero subindex signals the classical expression for the Casimirs.
These results fully agree with those obtained by other authors\footnote{A
word of caution is in order
concerning the normalization of the
$\F_k$. As mentioned before, it is related to the classical value of the
coupling $\tau_0$ through $\Lambda$.  For example in the case of
$SU(2)$ , if we want to have $\tau_0 = (i/\pi)(2\log{2} - 3)$,
in agreement with \cite{klemm}, we have to multiply every $\F_k$ in
(\ref{elresul}) with a scale factor $1/16^k$.} \cite{klemm, dhoker}.

\noindent
Concerning the following corrections, $\F_2$ and $\F_3$, the previous
formulae give
\[ \F_2^{SU(2)} = {5\ov 2^{8}\Delta_0^3} ~~~ , ~~~
\F_2^{SU(3)} = {3^2 u_0\ov 2^4 \Delta_0^3} (17 u_0^3 + 189
v_0^2) ~, \]
\beqas
\F_2^{SU(4)} & = & 2 [8 u_0^8 (20 u_0^3 - 373 v_0^2) 
+ 16 u_0^6 w_0 (206 u_0^3 - 2389 v_0^2) + 64 u_0^4 w_0^2 (464 u_0^3
- 2399 v_0^2) \\
& & + 2 u_0^5 (9207 v_0^4 + 71168 w_0^3) - 63 u_0^2 v_0^2 
(567 v_0^4 + 6400 w_0^3) + 16 u_0^3 w_0 (4977 v_0^4 \\ 
& & + 22016 w_0^3) + 36 v_0^2 w_0 (2403 v_0^4 - 23296 w_0^3) + 
32 u_0 w_0^2 (1323 v_0^4 + 11008 w_0^3] \Delta_0^{-3} ~,
\eeqas
\[ \F_3^{SU(2)} = {3\ov 2^{8}\Delta_0^5} ~~~ , ~~~
\F_3^{SU(3)} = {u_0\ov 2^4 \Delta_0^5} (3080 u_0^6 + 
119529 u_0^3 v_0^2 + 248589 v_0^4) ~, \]
which also agree\footnote{Note that the 
result for $\F_3^{SU(3)}$ in Ref.\cite{klemm} has a wrong global factor 
equal to $27$.} with those in Refs.\cite{klemm,dhoker,MNW}.
Despite these coincidences, we believe it would be useful to carry out
an exhaustive comparison with different methods, as well as a generalization to
other Lie algebras.

\section{Soft Supersymmetry Breaking with Higher  Casimir Operators}
\setcounter{equation}{0}
\indent

The prepotential of the Seiberg-Witten solution in the
Toda-Whitham framework depends on infinitely many slow times $T_n$, and
for $n=1, \dots, N-1$ we can find explicit expressions for its first and second
derivatives. In this section, we will interpret these slow times as
parameters of a nonsupersymmetric family of theories, by promoting them to
spurion
superfields in the spirit of \cite{soft,moresoft,luisIyII}. The higher order
slow times, as
(\ref{lasecudef}) shows, are dual to the $\H_{m}$, which are homogeneous
combinations of the Casimir operators of the group. This means
that we will be able to parametrize soft supersymmetry breaking terms induced
by
all the Casimirs of the group, and not just the quadratic one (the only case
considered
in \cite{moresoft}). This also generalizes to the ${\cal N}=0$ case the family
of ${\cal N}=1$ supersymmetry breaking terms considered for instance in
\cite{ad}. As the dependence on the slow
times as spurion superfields is encoded in the  prepotential, we will be able
to
obtain the exact effective potential of the theory, generalizing in this way
the results
of \cite{moresoft}.

\subsection{Properties of the prepotential under duality transformations}
To understand the softly broken model it is useful to derive
first
the properties of the prepotential and its derivatives under a duality
transformation in the effective theory, following the strategy of \cite{soft,
moresoft}. In the case of $SU(r+1)$ Yang-Mills theory,
the duality group is the symplectic group ${\rm Sp} (2r, {\bf Z})$. An element
of this
group has the structure
\be
\Gamma = \bemat{cc} A & B \\ C & D \enmat
\label{symp}
\ee
where the $r \times r$ matrices $A$, $B$, $C$, $D$ satisfy:
\be
A^t D-C^t B = {\bf 1}, \,\,\,\,\,\ A^tC= C^t A, \,\,\,\,\,\ B^tD= D^t B.
\label{conds}
\ee
In what follows it will be convenient to define the spurion variables $s_n$ as
\be
s_1=-i \log \Lambda, \,\,\,\,\ s_n= -i \hat T_n, \,\ n=2, \dots, r.
\label{spurions}
\ee
We take as our independent coordinates in the prepotential $\alpha^i, s_n$. The
dual
spurions are defined by
\be
s_{D,n}={\partial {\cal F} \over \partial s_n},
\label{timedual}
\ee
and we introduce a generalized $(2r) \times (2r)$ matrix of couplings as
follows:
\be
\tau_{ij}= {\partial^2 {\cal F}\over \partial \alpha^i \partial \alpha^j},
\quad
\tau_i{^n}= {\partial^2 {\cal F}\over \partial \alpha^i  \partial  s_n}
\quad
\tau^{mn}= {\partial^2 {\cal F}\over \partial s_m \partial s_n},
\label{coupl}
\ee
where $i,j,m,n=1, \dots, r$.

The symplectic group acts on the $\alpha^i$, $\alpha_{D,i}$ variables as
$v \rightarrow \Gamma v$, where $v^t =(\alpha_{D,i}, \alpha^i)$,
\be
\bemat{c}
\alpha^\Gamma_{D,i}\\  \rule{0mm}{6mm}
\alpha^{\Gamma\, i} \\
\enmat
=
\bemat{c}
{A_i}^k \alpha_{D,k} + B_{ik} \alpha^k \\ \rule{0mm}{6mm}
C^{ik}\alpha_{D,k} + {D^i}_k \alpha^k  \\
\enmat ~.
\ee
The spurion variables, coming from the slow times
of the Toda-Whitham hierarchy, parametrize deformations of the Seiberg-Witten
differential,
as we have seen in (\ref{alpha}), therefore they are invariant under the
duality transformations
(which are symplectic transformations of the homology cycles of the curve):
$s^\Gamma_m = s_m$. The
jacobian matrix of the change of coordinates is then given by
\be
\bemat{cc}
\pder{\alpha^{\Gamma\,i}}{\alpha^j} & \pder{\alpha^{\Gamma\, i}}{s_n}
\\ \rule{0mm}{7mm}
\pder{s^\Gamma_m}{\alpha^j} & \pder{s^\Gamma_m}{s_n} \\
\enmat
=
\bemat{cc}
C^{ik}\tau_{kj} + D^i{_j} & C^{ik}\tau_{kn} \\ \rule{0mm}{7mm}
0 & \delta^n_{m}
\enmat,
\label{jaco}
\ee
and we obtain the transformation laws for the operators
\beqa
\pder{~}{\alpha^{\Gamma\, i}} &=& \pder{\alpha^j}{\alpha^{\Gamma\, i}}
\pder{~}{\alpha^j} + \pder{s_m}{\alpha^{\Gamma\,i}}\pder{~}{s_m} \cr
&=& (C\tau + D)^{-1\,j}{_i} \pder{~}{\alpha^j}, \nonumber \\
&\rule{0mm}{7mm} &
\cr
\pder{~}{s_m^\Gamma} &=& \pder{\alpha^i}{s_m^\Gamma}\pder{~}{\alpha^i}
+ \pder{s_n}{s_m^\Gamma} \pder{~}{s_n}\cr
&=&
\pder{~}{s_m} - (C\tau + D)^{-1\,i}{_j} C^{jk}\tau_k{^m}\pder{~}{\alpha^i}.
\label{transder}
\eeqa
The transformation law for the prepotential has been found in
\cite{matone,cobi} and reads
\be
\F^{\Gamma} =
\F + {1\ov 2} \alpha^i(D^T B)_{ij} \alpha^j +
  {1\ov 2} \alpha_{D\,i}(C^T A)^{ij} \alpha_{D,j} +
      \alpha^i (B^T C)_i{^j} \alpha_{D,j}.
\label{gprepot}
\ee
Using (\ref{transder}), (\ref{gprepot}) and (\ref{conds}) one easily finds that
the dual times are
invariant under duality transformations:
\be
s^{\Gamma\,m}_D =  \pder{\F^\Gamma}{s_m^\Gamma} =  \pder{\F}{s_m}
= s^m_D
\label{dualinv}
\ee
and that the second derivatives of the prepotential transform as follows
\beqa
\tau^\Gamma_{ij} & = & (A\tau + B)(C\tau+ D)^{-1}_{ij} ~, \nonumber \\
{\tau^\Gamma}_i{^m} & = & \left[ (C\tau + D)^{-1}\right]^j{_i}
{}~\tau_j{^m} ~, \label{transtaumn} \\
\tau^{\Gamma\,mn} & = & \tau^{mn} - \tau^m{_i}\left[(C\tau + D)^{-1}
C \right]^{ij} \tau_j{^n} ~. \nonumber
\eeqa
Notice that the matrix $(C\tau + D)^{-1} C$ is symmetric (\cite{rauch}, p.
91).

Using (\ref{lasecu}) one can find explicit expressions for the dual spurions
and couplings
in terms of the gauge-invariant functions ${\cal H}_{n,m}$ and quantities
associated to the
hyperelliptic curve:
\beqa
s_D^1 & = & {\beta\ov 2\pi } \biggl[ \H_2 +i \sum_{m \ge 2} m s_m \H_{m+1}
- \sum_{m,n\ge 2} m s_m s_n \H_{m+1,n+1} \biggr] ~, \nonumber \\
s_D^n & = & {\beta\ov 2\pi  n} \biggl[ \H_{n+1} + i \sum_{m\geq 2} m s_m
\H_{m+1,n+1}\biggr] ~, \nonumber \\
\tau^1{_i}  & = & {\beta\ov 2\pi}\biggl[ \pder{\H_{2}}{a^i} + i\sum_{n\ge
2} s_n \pder{\H_{n+1}}{a^i} \biggr] ~, \nonumber \\
\tau^n{_i} & = & {\beta \ov 2\pi n} \pder{\H_{n+1}}{a^i} ~, \label{expu}\\
\tau^{11} & = & -2\tau^1{_i} \tau^1{_j} \d_{\tau_{ij}}
\log\Theta_E(0|\tau) ~, \nonumber \\
\tau^{1n} & = & -2 \tau^1{_i} \tau^n{_j}
\d_{\tau_{ij}}\log\Theta_E(0|\tau) ~, \nonumber \\
\tau^{nm} & = & {\beta\ov 2\pi i}\H_{m+1,n+1} -2 \tau^n{_i}\tau^m{_j}
\d_{\tau_{ij}}\log\Theta_E(0|\tau) ~. \nonumber
\eeqa
with $n,m \geq 2$.
Notice that, when the spurions $s_m$ are zero, we recover for the
variable $s_1$ the results of \cite{moresoft}.
One can check
that the explicit expressions for the couplings and the dual spurions obtained
in (\ref{expu}) satisfy the
transformation properties given in (\ref{transtaumn}).
The invariance of the dual times $s_{D,n}$, as expressed in (\ref{dualinv}),
is consistent with the fact that they depend  on
 $s_m$ and ${\cal H}_{m+1,n+1}$, which are duality invariant. To verify
the transformation
properties of $\tau^n{_i}$, we can now appeal to the transformation properties
of
the
derivatives of the
Casimirs,
\be
{\partial {\cal H}_{n+1} \over \partial \alpha^i} \rightarrow
[(C\tau+D)^{-1}]^j{_i}{\partial {\cal H}_{n+1} \over \partial \alpha^j},
\ee
which is again a consequence of the duality invariance of the $\H_{n+1}$.
Finally,
to obtain the transformation law of $\tau^{mn}$,
we need the behaviour of the theta function under
the symplectic transformation (\ref{symp}). The arguments $\xi$, $\tau$ of the
theta function change as follows:
\bea
\tau & \rightarrow & \tau^{\Gamma} = (A \tau + B) (C \tau + D)^{-1}, \nonumber
\\
\xi & \rightarrow & \xi^{\Gamma} = [ (C\tau + D)^{-1}]^t \xi,
\eea
and the characteristics (understood as row vectors) transform as
\bea
\alpha &\rightarrow& \alpha ^{\Gamma}= D\alpha-C\beta +{1 \over 2} {\rm
diag}(CD^t), \\
\beta  &\rightarrow& \beta ^{\Gamma}= -B\alpha+A\beta +{1 \over 2} {\rm
diag}
(AB^t).
\label{chartrans}
\eea
The transformation law for the theta function is then given by \cite{rauch}
\be
\Theta [\alpha^{\Gamma}, \beta^{\Gamma}] (\tau^{\Gamma}|
\xi^{\Gamma})=
K \exp \bigl[ \pi i \xi^t (C\tau + D)^{-1} C \xi \bigr] \Theta [
\alpha , \beta](\tau|\xi),
\label{thetatrans}
\ee
where $K$ is given by
\be
K= {\rm e}^{i \phi} ({\rm det} (C\tau + D))^{1/2}
\ee
and $\phi$ is a $\xi$-independent phase that will cancel in the logarithmic
derivative. Using (\ref{thetatrans}) we see that, under a symplectic
transformation,
$\partial_{\tau_{ij}}\log \Theta(0|\tau)$ gets shifted by a term of the form
\be
{1 \over 2} (C\tau+D)^i{_k}  (C\tau+D)^j{_l} [(C\tau+D)^{-1}C]^{kl},
\label{shift}
\ee
and in this way we recover (\ref{transtaumn}) directly from (\ref{expu}).

\subsection{The Microscopic Lagrangian}
As anticipated above, to break ${\cal N} = 2$ supersymmetry down to ${\cal N} =
0$,
we promote the variables $s_n$ to ${\cal N} = 2$ vector superfields $S_n$, and
then freeze the scalar and auxiliary components to constant vacuum expectation
values. Indeed,
we are really interested in non-supersymmetric deformations of pure $SU(r+1)$
Yang--Mills theory that preserve the nice holomorphic properties of the
Seiberg--Witten solution. Thus, for all $S_n, n \geq 2$, we will set the
corresponding scalar components $s_n$ to zero, while keeping the
top components $D_n$ and $F_n$ as supersymmetry breaking parameters. A similar
scenario was considered in \cite{luisIyII}, where the bare masses of the
hypermultiplets were also promoted to ${\cal N}=2$ superfields, and
non-supersymmetric deformations of the {\it massless} theory were studied by
turning on the
auxiliary components of the mass spurion superfield while setting the scalar
components
(the bare masses of the original model) equal to zero.

The couplings of the
spurions $S_n$ are encoded in the holomorphic dependence on the slow times.
In terms of ${\cal N} = 1$ superfields we have,
\be
S \equiv S_1 = s_1 + \theta^2 F_1 \;\;\; , \;\;\; V_s \equiv V_1 = \med
D_1 \theta^2 \bar\theta^2 ~,
\label{vevs1}
\ee
\be
S_n = \theta^2 F_n \;\;\; , \;\;\; V_n = \med D_n \theta^2 \bar\theta^2 ~,
\,\,\ n\ge 2
\label{vevs2}
\ee
where $s_1$ is related to the dynamical scale of the theory, $\Lambda
= e^{is_1}$.  Although the prepotential of the Whitham hierarchy is defined for
the low-energy theory, it is important to know  what is the microscopic,
non-supersymmetric
theory whose effective action is encoded in the prepotential $\F (\alpha_i,
S_n)$ (where
the $S_n$ are given in (\ref{vevs1})--(\ref{vevs2})). When we only consider the
first
slow time, $S_1$, we can identify it with the classical gauge coupling in the
bare
Lagrangian. In
fact, the RG equation for the $SU(r+1)$ theory gives a relation between
$\Lambda$
and the coupling $\tau = \frac{\theta}{2\pi} + \frac{4\pi{i}}{g^2}$,
$\Lambda^{2(r+1)} \sim e^{2i\pi\tau}$. We then see that the scalar component
of $S_1$ can be written as $s_1 = \pi\tau/(r+1)$, and the spurion superfield
appears in the classical prepotential as follows \cite{luisIyII,hsu}
\be
\F = \frac{r+1}{2\pi} S_1 \, \tr A^2 = \frac{r+1}{\pi} S_1 \H_2 ~.
\label{class1}
\ee
The nonsupersymmetric microscopic Lagrangian is then obtained
from (\ref{class1}) by turning on the scalar and auxiliary components of $S_1$.

Let us now analyze what happens when the rest of the slow times are turned on.
We can expand the prepotential
around $S_2= \cdots=S_{r}=0$, but then, in the supersymmetric case, we should
take into account the higher derivatives of the prepotential to obtain the
terms of order ${\cal O}(S^3)$ in this expansion. If we consider however the
case of softly broken supersymmetry, where the $S_{n \ge 2}$ have the form 
(\ref{vevs1})--(\ref{vevs2}), then the terms with more that two
powers of the $S_{n \ge 2}$ will not give any term in the Lagrangian, because
they involve too many $\theta$'s and the integral in superspace will vanish.
Notice that this is a general fact that holds for both the macroscopic and the
microscopic Lagrangian: as we only know the first and second derivatives of the
prepotential, we will be able to write the complete Lagrangian only when we
restrict ourselves to the spurion configurations (\ref{vevs1})--(\ref{vevs2}). 
Also notice
that, as the derivatives of the prepotential are evaluated at $S_2=\cdots
=S_r=0$, they
are obtained by evaluating the right hand side of (2.12) in the original
Seiberg-Witten solution.

To write the microscopic Lagrangian, we have to consider the classical limit of
the expressions
(\ref{expu}) evaluated at $s_2=\cdots=s_{r}=0$. From the expressions we have
obtained
in section 3, it is easy to see that:
\be
\pder{\H_{r+1}}{a^i}\pder{\H_{r+1}}{a^j}{1\ov i\pi}
\d_{\tau_{ij}}\log\Theta_E(0|\tau) \sim \Lambda^{2r} {\cal O} \biggl(
{\Lambda^2 \over Z_{\alpha_+} ^2} \biggr),
\label{thetavan}
\ee
then the terms involving the theta function vanish in the semiclassical limit
$(\Lambda/ Z_{\alpha_+}) \rightarrow 0$ (for all the positive roots
$\alpha_+$). This is also the expected
behaviour of these terms in the twisted theory: they are contact terms that
should vanish at tree level \cite{mw}. The microscopic Lagrangian we are
considering
is then given by the following deformed, classical prepotential
\be
\F = \frac{r+1}{\pi} \sum_{n=1}^r \frac{1}{n} S_n \H_{n+1} + {r+1 \over 2\pi i
} \sum_{n,m\ge2}
S_nS_m \H_{n+1, m+1} ~,
\label{class2}
\ee
where the spurions $S_n$, $n\ge1$ are of the form (\ref{vevs1})--(\ref{vevs2}).
We then see that we are studying nonsupersymmetric deformations
of the ${\cal N}=2$ theory induced by the higher Casimirs. We will be mainly
interested in the
``reduced" prepotential,
\be
\F^{\rm red} = \F- {r+1 \over 2\pi i } \sum_{n,m\ge2}
S_n S_m \H_{n+1, m+1},
\label{red}
\ee
as in this case the perturbations are linear in the
higher order slow times
(at the classical level) and only involve the Casimirs $\H_{n+1}$. Notice that
the
second derivative
of the reduced, quantum prepotential (when the higher order times are set to
zero) is then given by
\be
\ppder{\F^{\rm red}}{S_m}{S_n}\biggl|_{S_2= \cdots= S_{r}=0} =
{\beta^2\ov 2\pi i m n } \pder{\H_{m+1}}{a^i}\pder{\H_{n+1}}{a^j}{1\ov
i\pi} \d_{\tau_{ij}}\log\Theta_E(0|\tau)~,~~~(m,n\geq  2)
\label{reder}
\ee
an expression which is invariant under the semiclassical monodromy and vanishes
semiclassically. This reduced prepotential is in fact the relevant one for
Donaldson-Witten theory, and the second derivatives (\ref{reder}) are
essentially the
contact terms for higher Casimirs found in \cite{losev}.  From now on we will
consider
the reduced prepotential and omit the superscript.

If we expand (\ref{red}) in superspace, we find the microscopic Lagrangian
\be
{\cal L} = L_{kin} + L_{int} ~,
\label{lagra}
\ee
where,
\beqa
L_{kin} & = & {1\ov 4\pi} {\rm Im} \bigl[ (\nabla_\mu \phi)^\dagger_a
(\nabla^\mu \F)^a + i(\nabla_\mu\psi)^\dagger_a \bsigma^\mu \psi^b {\F^a}_b
- i \F_{ab} \lambda^a\sigma^\mu (\nabla_\mu\blam)^b \nonumber \\
& - & {1\ov 4}\F_{ab}\, ({F^a}_{\mu\nu} F^{b\,\mu\nu} + i
{F^a}_{\mu\nu} \tilde F^{b\,\mu\nu}) \bigr] ~,
\label{lkin}
\eeqa
\beqa
L_{int} & = & {1\ov 4\pi} {\rm Im} \bigl[ \F_{AB} F^A {F^*}^B -
{1\ov 2} \F_{abC} \left( (\psi^a\psi^b) {F^*}^C + (\lambda^a\lambda^b)
F^C + i\sqrt{2} (\psi^a \lambda^b) D^C \right) \nonumber \\
& + & {1\ov 2} \F_{AB} D^A D^B + ig (\phi^*_a {f^a}_{bc} D_b \F^c
+ \sqrt{2} \left[(\phi^*\lambda)_a {\F^a}_{b}\psi^b -
(\bpsi\blam)_a\F^a \right]) \bigr] ~.
\label{lint}
\eeqa
In (\ref{lkin})--(\ref{lint}), $\lambda$, $\psi$ are the gluinos and
$\phi$ is the scalar component of the ${\cal N} = 2$ vector superfield.
 The $f^a_{~bc}$ are the
structure constants of the Lie algebra.
The indices $a,b,c,...$ belong to the adjoint representation of $SU(N)$,
and are raised and lowered with the invariant metric.
 $A,B,...$ run over both the indices in the adjoint,
\,$a,b,...$, and over those of the slow times,\, $n,m,...$~.
Since all spurions corresponding to higher Casimirs are purely auxiliary
superfields, (\ref{vevs2}), the Lagrangian can be easily presented
in a more transparent form. To this end we first integrate the
auxiliary fields out
\be
D^a = - b_{\cl}^{-1~ac} \left( b^{\cl}{_c}{^m} D_m +
\Re (g \, a^*_b {f^b}_{ca}\F^a )
\right) ~,
\label{auxil1}
\ee
\be
F^a = - b_{\cl}^{-1~ac} \, b^{\cl}{_c}{^m} F_m ~.
\label{auxil2}
\ee
In these equations, we have introduced the classical matrix of couplings
$b_{AB}^{\cl}$,
\be
b^{\cl} = {1 \over 4 \pi} {\rm Im} \,\tau ^{\cl} ~,
\label{imclass}
\ee
which are given by the derivatives of $\F^{\rm red}$ with
respect to the lower components of the vector superfields. We obtain the
following expressions for them,
\be
\tau_{ab}^{\cl} = \tau\,\delta_{ab} ~,
\label{babcl}
\ee
\be
\tau^{\cl}{_a}{^m} = \frac{N}{\pi{i}m} \pder{\H^{\cl}_{m+1}}{\phi^a} =
\frac{N}{\pi{i}m} \tr\,(\phi^m{T_a}) + \ldots  ~,
\label{bamcl}
\ee
\be
\tau^{mn}_{\cl} = 0 ~,
\label{bmncl}
\ee
where the dots (in eq.(\ref{bamcl})) denote the derivative with respect
to $\phi^a$ of lower order Casimir operators.
Inserting back (\ref{auxil1})--(\ref{auxil2}) in (\ref{lagra}), we find
\beqa
{\cal L} & = & {\cal L}_{N=2}  - B^{mn}_{\cl} \left( F_m F^*_{n} +
{1\ov 2} D_m D_n \right) + f^e_{~bc} b^{\cl}{_a}{^m} {b^{\cl}}^{-1}_{ae} D_m
\phi^{b} {\bar\phi}^c
\nonumber \\
& + & {1\ov 8\pi}{\rm Im}
\frac{\partial\tau^{\cl}{_b}{^m}}{\partial\phi^a}\Bigg[
(\psi^a\psi^b)
F^*_m + (\lambda^a\lambda^b) F_m + i\sqrt{2}(\lambda^a\psi^b) D_m\Bigg] ~,
\label{barelag}
\eeqa
where $B^{mn}_{\cl}$ is the classical value of the duality invariant quantity
\be
B^{mn} = b_{a}{^m}{b}^{-1\,ab}b_b{^n} - b^{mn} ~.
\label{moninv}
\ee
The dilaton spurion gives mass to the gauginos of the ${\cal N} = 2$
vector multiplet and to the imaginary part of the Higgs field $\phi$.
The spurions corresponding to higher Casimirs, on the other hand,
give couplings between the Higgs field and the
gauginos. Finally,
note that the bare Lagrangian (\ref{barelag}) is not CP invariant,
since $\tau$ and $F_m$ are arbitrary complex parameters. Thus, the
corresponding low energy effective Lagrangian on which we will focus
from now on, is not in general CP invariant. Notice that the spurion
superfields
$S_n$ have dimensions $1-n$, therefore the supersymmetry breaking parameters
$F_n$, $D_n$ have dimension $2-n$. For $n>2$, they will give nonrenormalizable
({\it i.e.}
irrelevant) interactions in the microscopic Lagrangian. This does not mean that
the resulting perturbations do not change the low-energy structure of the
theory: the
operators we are considering can be dangerously irrelevant operators, as in the
related theory analyzed in \cite{seiberg}, and in this case
they will affect the infrared physics.

\subsection{The Effective Potential}
As the prepotential has an analytic dependence on the spurion superfields,
the effective Lagrangian up to two derivatives and four fermion terms for
the ${\cal N} = 0$ theory described by (\ref{lagra}) is given by the exact
Seiberg-Witten solution once the spurion superfields are taken into account.
This gives the exact effective potential at leading order and the vacuum
structure can be determined. The computation of the effective potential
and the condensates is very similar to the one in \cite{moresoft}.
If we are near a submanifold of the moduli space of vacua where $n_H$
hypermultiplets
become massless, the full Lagrangian contains the vector multiplet contribution
and
the hypermultiplet contribution involving (in ${\cal N}=1$ language) two
chiral superfields
$H_a$, $\tilde H_a$, $a=1, \dots, n_H$. We choose an appropriate
duality frame whose variables
will be generically denoted by $a_i$. We will denote the charge of the $a$th
hypermultiplet with
respect to the $i$th $U(1)$ factor by $n^a_i$, where $a=1, \dots, n_H$. In this
section, the indices $m,n=1, \dots, r$ label the spurion variables, $s_m$,
and the indices $i,j=1, \dots, r$
 the period variables, $a^i$.
The matrix of couplings appearing in the effective potential is given
by
\be
b={1 \over 4 \pi} {\rm Im}\,\tau~.
\label{imcop}
\ee
If we now define the quantities
\beqa
(n^a,n^b) & = & n^a_i b^{-1\,ij} n^b_j ~, \nonumber \\
(n^a,b^m) & = & n^a_i b^{-1\,ij} b_j{^m} ~,
\label{quantis}
\eeqa
the effective potential can be written as
\beqa
V & = & B^{mn} \left( F_m F^*_{n} + {1\ov 2} D_m D_n \right)
+  (n^a,b_m) D^m \left( |h_a|^2- |\tilde h_a|^2 \right) \nonumber \\
& + & \sqrt{2}\,(n^a,b^m) \left(  F_m\tilde h_a h_a +
\bar F_m\bar h_a\bar{\tilde h}_a \right)
+ 2 (n^a,n^b)( h_a{\tilde h}_a \bar{h}_b \bar{\tilde h}_b) \\
& + & {1\ov 2} (n^a,n^b)(|h_a|^2-|\tilde h_a|^2)(|h_b|^2-|\tilde h_b|^2)
+ 2 |n^a \cdotsh a|^2(|h_a|^2 + |\tilde h_a|^2) ~, \nonumber
\label{final}
\eeqa
where $n^a \cdotsh a= \sum_i n^a_i a^i$, and $h_a$ ($\tilde h_a$) is the scalar
component of $H_a$ ($\tilde H_a$). This expression is identical to the
one derived in \cite{moresoft}, with the only difference that
we have now $r$ spurion superfields associated to the different Casimirs of
the group. We can then adapt the results derived there to this context,
where we also set $D_m=0$, these being no real restriction since $F_n$ and
$D_n$ transform as doublets under $SU(2)_R$. 
To obtain the
values of the condensates, we minimize $V$ with respect to $h_a$, $\tilde h_a$.
One finds that $|h_a| = |\tilde h_a|$.
It is convenient to fix the gauge in the $U(1)^r$ factors in such a way that
\be
h_a = \rho_a  ~~,~~~
\tilde h_a = \rho_a e^{i\beta_a}
\label{gaugefix}
\ee
If the charge vectors $n^a$ are
linearly independent, the non trivial condensates are given by
\be
|n^a \cdotsh a|^2 + \sum_b(n^a,n^b)\rho_b^2
e^{i(\beta_b-\beta_a)}
 + {1\ov \sqrt{2}}(n^a,b^m) F_m  e^{-i \beta_a } = 0,
\label{condensados}
\ee
 The effective potential is then given by
\be
V= B^{mn}F_mF^*_n - 2
\sum_{ab}(n^a,n^b)\rho_a^2\rho_b^2
\cos(\beta_a- \beta_b)
\label{potefftres}
\ee
To make use of the previous equations one needs explicit 
knowledge  of the values of the couplings as functions over the moduli space
 $\tau(a)$. This is achieved by means of eqs.(\ref{expu}) where the 
terms on the right hand side can be computed from the original Seiberg-Witten
solution. We shall see two examples in the following section.

\section{Analysis of SU($3$)}
\setcounter{equation}{0}
\indent

$N=2$ supersymmetric Yang-Mills theory with gauge group SU($3$) has been
analyzed in detail in \cite{klemm,ad}.
As it is well-known, there are two sets of distinguished points in the moduli
space of the theory. The three ${\bf Z}_2$ vacua are located at
$u^3=(3\Lambda^2)^3$, $v=0$, and give rise to the
${\cal N}=1$ vacua when the theory is perturbed with a mass term of the form
${\rm Tr} \Phi^2$ (we denote $u_2=u$, $u_3=v$).
Then we have the two ${\bf Z}_3$ vacua, located at $u=0$ and $v=\pm 2
\Lambda^2$. They are also known as the Argyres-Douglas (AD) points, and there
are three mutually nonlocal BPS states becoming massless at this point. The
low-energy theory there is an ${\cal N}=2$ superconformal theory and the two
$U(1)$ factors are decoupled.

In this section we will briefly examine the softly broken theory near these
vacua. This will
also illustrate the structure of the formalism we have been using in the strong
coupling regime, in particular the use of theta functions. We will set
$\Lambda=1$, as in \cite{ad,ds,ITEP}.

\subsection{The ${\bf Z}_2$ vacua}
In this subsection we study the soft breaking of the theory near the ${\cal
N}=1$
points, which have been studied in detail in
\cite{ds}. To evaluate the second derivatives of the prepotential, we need the
values of
the periods of the hyperelliptic curve and the structure of the gauge
couplings. We will focus on the ${\cal N}=1$ point where $N-1$ magnetic
monopoles
become massless (corresponding in $SU(3)$ to the point $u=3 \Lambda^2$, $v=0$).
The values of the quantities at the other points can be obtained using the
${\bf Z}_N$ symmetry in the moduli space. The eigenvalues for the field $\phi$
are given by
\be
\phi_n = 2 \cos {\pi (n-{1 \over 2}) \over N}, \,\,\ n=1, \cdots, N.
\label{eigen}
\ee
The derivatives of the dual variables satisfy
\be
\sum_{j=1}^r {\partial a_{D, j} \over \partial \phi_n} \sin {\pi
j l  \over N} =
i \cos {\pi l (n-{1 \over 2}) \over N},
\label{ortonor}
\ee
and from this relation one can easily derive the general result
\be
 {1\ov n}{\partial {\rm Tr} \, \phi^n  \over \partial a_{D,j}}= -2i
\sum_{l=0}^{[n/2-1]}
{n-1 \choose l}
\sin { \pi j (n-2l-1) \over N}
\ee
for $SU(N)$. In $SU(3)$, $u = \med \Tr\, \phi^2$ and $v = {1\ov 3}\Tr
\,\phi^3$, so that
\be
{\partial u  \over \partial a_{D,j}}=-2i \sin { \pi j \over N}~~, ~~~~~~~~~
{\partial v  \over \partial a_{D,j}}=-2i \sin { 2\pi j \over N}.
\label{derivs}
\ee
The gauge couplings near the ${\cal N}=1$ point have the
structure
\be
\tau^D_{ij}= {1 \over 2\pi i } \log \left( {a_{D,i}\over \Lambda_i}  \right)
\delta_{ij} + (1-\delta_{ij})\tau_{ij}^{\rm off}(0) + {\cal O}(a_{Di}),
\ee
where $\Lambda_i/\Lambda \sim \sin (\pi i /N)$, and $\tau_{ij}^{\rm off}(0)$,
$i \not=j$ are the values of the off-diagonal entries of the coupling constant
at the ${\cal N}=1$ point $a_{Di}=0$.
For $SU(3)$, $\tau_{12}={i \over \pi} \log 2$ \cite{klemm}. For $SU(N)$, the
$\tau_{ij}^{\rm off}(0)$ can be obtained from the results on the scaling
trajectory in section 5 of \cite{ds}.
To compute the theta function in magnetic variables, we have to take into
account the
change of the ``electric" characteristics under the symplectic transformation
\be
\Omega= \bemat{cc} 0& {\bf 1} \\
-{\bf 1} & 0 \enmat
\label{magdual}
\ee
to the magnetic variables $a_{D\,i}$.
Using (\ref{chartrans}) we find
\be
\vec \alpha= (1/2, \dots, 1/2), \,\,\,\,\,\ \vec \beta=(0, \dots, 0).
\label{dualchar}
\ee
One can then obtain, at the ${\cal N}=1$ point of $SU(3)$,
\be
{1\ov i\pi} \d_{\tau_{ij}} \log\Theta_D(0|\tau^D)
 =
{1\ov 4}\delta_{ij} - {1\ov 12}(1-\delta_{ij}).
\label{nonetheta}
\ee
Using  (\ref{derivs}) and (\ref{nonetheta}), it is easy to check the relation
(\ref{qucas}) for the $\Lambda$ derivatives of $u$ and $v$ at this ${\cal N}=1$
point. For $SU(N)$, the diagonal part of the matrix (\ref{nonetheta}) is still
of the form
$(1/4)\delta_{ij}$, but the off-diagonal part is more involved and one needs
the values of the couplings
$\tau_{ij}^{\rm off}(0)$.

With this information we can already discuss the structure of the
condensates at the
${\cal N}=1$ points, following the discussion in section 4 of \cite{moresoft}.
At the point where $N-1$ monopoles become massless, there is a simplectic basis
for the hyperelliptic curve, such that the magnetic
charge vectors are given by $n^a_j=\delta^a_j$, and the equation
(\ref{condensados}) becomes
\be
\rho^2_i=-\sum_{j}{\rm e}^{i(\beta_j-\beta_i)} b_{ij}|a_j|^2 -{{\rm
e}^{-i\beta_i}
\over {\sqrt 2}} \sum_{n=1,2} F_n b^n{_i}
\label{nonecond}
\ee
At the ${\cal N}=1$ point, $a_{D,i}=0$ and the condensates are essentially
given by
the absolute value of the order
parameters $b^n{_i} = (N/4\pi^2) {\rm Im}(\d H_{n+1}/\d a^i)$, which in the
case of
$SU(3)$ can be obtained  from (\ref{derivs})
\be
b^1{_i}=-{3 \over 2\pi^2} \sin{\pi i \over 3}~~,~~~~~~~~~  b^2{_i}= -{3 \over
4\pi^2} \sin{2\pi i \over 3} .
\label{order parameters}
\ee
Hence
\be
\rho^2_1 = \sqrt{3\ov 2} \,{3\ov 4\pi^2} \,|F_1 + \med F_2| ~~~,~~~~~~~~~
\rho^2_2 = \sqrt{3\ov 2} \, {3\ov 4\pi^2} \,|F_1 - \med F_2|~. \nonumber
\ee
and we see that the soft breaking induced by the quadratic and
cubic Casimirs gives rise to monopole condensation in both $U(1)$ factors,
although the condensates (and therefore the string tension) are
bigger for the soft breaking coming from $u$ (for equal values of the
supersymmetry breaking
parameters $F_1$, $F_2$). In the same way, the vacuum energy associated to
these condensates is
\beqa
 V_{\rm eff} &=& - b^{mn} F_m F^*_n \nonumber\\
&=& -{9\over 4\pi ^2} \left( |F_1|^2 + \med |F_2|^2\right)~.
\label{vacen}
\eeqa
As expected, the soft breaking associated to $u$ gives lower energy to the
vacuum.

\subsection{The $Z_3$ vacua}
Next we explore the behaviour near the Argyres-Douglas point at $v=2\Lambda^3$,
$u=0$.
It is convenient to use the parameters $\rho$ and $\epsilon$ introduced
in \cite{ad} and defined by
\be
u= 3 \epsilon^2 \rho, \,\,\,\,\  v-2\Lambda^3 = 2 \epsilon^3.
\label{rhoeps}
\ee
The three submanifolds $\rho^3=1$ correspond to three massless BPS states which
after
an appropriate symplectic transformation can be seen to be charged with respect
to only one of the $U(1)$ factors, with variables denoted by $a^1, a_{D,1}$.
Using the symplectic transformation of
\cite{moore2}, the charges of these states with respect to the $a^1, a_{D,1}$
are
$(n_e, n_m) = (-1,0)$, $(1,-1)$ and $(0,1)$,
{\it i.e.} we have one electron, one dyon, and one monopole. These submanifolds
come together
at the AD point, where we have a nontrivial superconformal field theory. The
two $U(1)$ 's are weakly coupled near the AD point, and the hyperelliptic
curve splits into
a small torus (corresponding to two mutually nonlocal periods $a^1, a_{D,1}$
which go to zero) and a big torus with
periods $a^2, a_{D,2}\sim \Lambda$. The small torus is given by the elliptic
curve
\be
w^2=z^3-3 \rho z -2,
\label{rhocurve}
\ee
and the meromorphic Seiberg-Witten differential degenerates on (\ref{rhocurve})
to
\be
\lambda_{SW} = {1 \over 2\pi} {\epsilon^{5/2} \over  \Lambda^{3/2} } w dz.
\label{swdif}
\ee
The matrix $\partial a^i/\partial u_j$ near the AD point reads, at leading
order \cite{moore2}:
\be
\bemat{cc} {\partial a^1 \over \partial u}\rule{0mm}{4mm} & {\partial a^1 \over
\partial v}
\\ \rule{0mm}{4mm}
{\partial a^2 \over \partial u} & {\partial a^2 \over \partial
v}\enmat =
\bemat{cc} { 2\epsilon^{1/2} \over \pi \Lambda^{3/2}} \eta & -{\epsilon^{-1/2}
\over 4\pi  \Lambda^{3/2}} \omega_{\rho} \\
{c \over \Lambda} & {d \over \Lambda^2} \enmat,
\label{adperiods}
\ee
where $\omega_\rho$ is the period of the elliptic curve (\ref{rhocurve})
corresponding
to $a_1$ (with
 ${\rm Im}(\omega_{\rho,D}/\omega_{\rho})>0$),  $\eta=\zeta(\omega_{\rho}/2)$
is the
value of the Weierstrass zeta function at the half-period, and $c,d$ are
nonzero constants (which can be obtained from the explicit computations
in \cite{ms,ky}).
For the dual variables we have similar expressions with $\omega_{\rho,D}$,
$\eta_D$,
$c_D$ and $d_D$. Using these expressions one can obtain the matrix of couplings
near the AD point \cite{moore2,ad,ky}
\beqa
\tau_{11} &= & \tau(\rho) + {\cal O}(\epsilon) ,\nonumber\\
\tau_{12} &= & -{2i \over c \Lambda^{1/2}} {\epsilon^{1/2}\over
\omega_{\rho}}+ {\cal O}(\epsilon^{3/2}), \nonumber\\
\tau_{22}&=& \omega + {\cal O}(\epsilon),
\label{couplings}
\eeqa
where $\omega= {\rm e}^{\pi i/3}$.

To analyze the theta function in these variables, we need the symplectic
transformation from
the electric variables to the variables appropriate for the large and the small
torus. We first compute the transformation of the characteristics under this
symplectic transformation. Using
(\ref{chartrans}) and the results in \cite{moore2} we find
\be
\vec \alpha= \vec\beta= (1/2,1/2).
\label{adchar}
\ee
We can already obtain the behaviour of the theta function as an expansion in
$\epsilon$:
\be
\Theta(0|\tau)  = -{1 \over \pi c \Lambda^{1/2}}  {\epsilon^{1/2}\over
\omega_{\rho}} \vartheta_1'(0|\tau(\rho)) \vartheta_1'(0|\omega) + {\cal
O}(\epsilon^{3/2}) ,
\label{theta}
\ee
where $\vartheta_1(\xi|\tau)$ is the Jacobi theta function with characteristic
$[1/2,1/2]$. Using
that
\be
{\vartheta_1'''(0|\tau)\over \vartheta_1'(0|\tau) } =-\pi^2 E_2(\tau),
\label{ident}
\ee
we find
\be
{1 \over i\pi} \partial_{\tau_{ij}}\log \Theta = \bemat{cc} {1\over 4} E_2
(\tau(\rho)) &
{c \Lambda^{1/2} \over 4\pi} \epsilon^{-1/2} \omega_{\rho} \\
{c \Lambda^{1/2} \over 4\pi} \epsilon^{-1/2} \omega_{\rho} & {1 \over 4}
E_2(\omega) \enmat.
\label{thetamat}
\ee
Again, using (\ref{adperiods}) and (\ref{thetamat}), one can check the relation
(\ref{qucas}) for $v$ (for $u$ one needs the explicit values of the constants
appearing in
the above expressions).

The analysis of the condensates near the AD point is difficult because one has
to take into account mutually
nonlocal degrees of freedom,
and there is not a Lagrangian description of this theory. In fact, one expects
that, in the softly
broken theory, a cusp singularity will appear in the effective potential near
the AD point, as it happens in
${\cal N}=2$
QCD with gauge group $SU(2)$ and one massive flavour \cite{luisIyII}. But we
can analyze the monopole condensates along the divisors $\rho^3=1$ and their
evolution
as we approach the AD point.  Near each of the submanifolds $\rho^3=1$ there is
a massless BPS state, and we
expect it to condense after breaking supersymmetry down to ${\cal N}=0$. These
condensates correspond to mutually nonlocal states, but we can assume, as in
\cite{moresoft, luisIyII}, that
these states do not interact and that the condensates are given by the equation
\be
\rho^2_k=-{1 \over (b^{-1})_{11} }|a_k|^2 - {{\rm e}^{-i\beta_k} \over {\sqrt
2} (b^{-1})_{11} } \sum_{n=1,2}F_n
(b^{-1})_{1j} b^n_{~j},
\label{rhoconds}
\ee
where $k=1,2,3$ and $a_k$ are the appropriate local coordinates for each of the
massless states ({\it i.e.} $a_k=a^1$, $a_{D,1}$, $a^1-a_{D,1}$). The equation
(\ref{rhoconds})
can be obtained from (\ref{condensados}) taking into account that the states
are only charged with respecto to the first $U(1)$ factor. The quantities
$(b^{-1})_{ij}$, $b^n{_j}$ should be also computed in the duality frame
dictated by the
$a_k$. For example, for $a^1$ we use the
``electric" period of the $\rho$ curve, $\omega_{\rho}$, and for
$a_{D,1}$ we use
the ``magnetic" period $\omega_{\rho,D}$. The approximation where the
mutually nonlocal states do not interact should be good far enough from the
AD point. These condensates give only a magnetic Higgs mechanism in one of
the $U(1)$ factors, and correspond to the
half-Higgsed vacua of \cite{ad}. Notice that one should perform a careful
numerical study
of the equations for the condensates and for the effective potential to know
if these partial
condensates  give the true vacua of the ${\cal N}=0$ theory. As we approach
the AD point, $\epsilon \rightarrow 0$, we see that the parameters for
condensation go to
zero for both the quadratic and the cubic Casimir:
\be
{\partial u\over \partial a^1}, \,\ {\partial v\over \partial a^1} \sim {\cal
O}(\epsilon^{1/2}),
\label{gozero}
\ee
and the mass gap associated to the condensates vanishes at the AD point, like
in the
${\cal N}=1$ breaking considered in \cite{ad}.

\vspace{1.5cm}


{\large\bf Acknowledgements}
\bigskip

We acknowledge comments from  J.L. Miramontes and  A. Morozov. We would also
like to thank  F. Zamora for useful suggestions on the manuscript and
  J.O. Madsen for computing assistance.
 M.M. would like to thank G. Moore for many useful discussions on integrable
systems
and ${\cal N}=2$ theories. The work of J.D.E. is supported by a fellowship of
the
Ministry of Education and Culture of Spain. The work of M.M. is supported by
DOE grant
DE-FG02-92ER40704. The work of J.M. was partially supported by DGCIYT under
contract
PB96-0960.


\end{document}